\documentclass{article}

\usepackage{arxiv}

\usepackage[utf8]{inputenc} % allow utf-8 input
\usepackage[T1]{fontenc}    % use 8-bit T1 fonts
\usepackage{hyperref}       % hyperlinks
\usepackage{url}            % simple URL typesetting
\usepackage{booktabs}       % professional-quality tables
\usepackage{amsfonts}       % blackboard math symbols
\usepackage{nicefrac}       % compact symbols for 1/2, etc.
\usepackage{microtype}      % microtypography
\usepackage{lipsum}
\usepackage{fancyhdr}       % header
\usepackage{graphicx}       % graphics
\graphicspath{{media/}}     % organize your images and other figures under media/ folder

\title{
 \textit{De novo} Design of Polymer Electrolytes with High Conductivity using GPT-based and Diffusion-based Generative Models }

\author{
  Zhenze Yang\textsuperscript{1,2}, Weike Ye\textsuperscript{†1}, Xiangyun Lei\textsuperscript{†1},   Daniel Schweigert\textsuperscript{1}, Ha-Kyung Kwon\textsuperscript{1}, Arash Khajeh\textsuperscript{1†*} \\
  \textsuperscript{1}Toyota Research Institute, \\
  4440 El Camino Real, Los Altos, California 94022, United States of America.\\
  \textsuperscript{2}Department of Materials Science and Engineering, \\
  Massachusetts Institute of Technology, \\
  77 Massachusetts Ave., Cambridge, Massachusetts 02139, United States of America \\
  \texttt{arash.khajeh@tri.global} \\
}

\begin{document}
\maketitle
\begin{abstract}
Solid polymer electrolytes hold significant promise as materials for next-generation batteries due to their superior safety performance, enhanced specific energy, and extended lifespans compared to liquid electrolytes. However, the material's low ionic conductivity impedes its commercialization, and the vast polymer space poses significant challenges for the screening and design. In this study, we assess the capabilities of generative artificial intelligence (AI) for the de novo design of polymer electrolytes. To optimize the generation, we compare different deep learning architectures, including both GPT-based and diffusion-based models, and benchmark the results with hyperparameter tuning. We further employ various evaluation metrics and full-atom molecular dynamics simulations to assess the performance of different generative model architectures and to validate the top candidates produced by each model. Out of only 45 candidates being tested, we discovered 17 polymers that achieve superior ionic conductivity better than any other polymers in our database, with some of them doubling the conductivity value. In addition, by adopting a pretraining and fine-tuning methodology, we significantly improve the efficacy of our generative models, achieving quicker convergence, enhanced performance with limited data, and greater diversity. Using the proposed method, we can easily generate a large number of novel, diverse, and valid polymers, with a chance of synthesizability, enabling us to identify promising candidates with markedly improved efficiency.
\end{abstract}

\section{Introduction}

Solid polymer electrolytes (SPEs) are promising candidates as next-generation lithium-ion battery materials, given their excellent safety, energy, and manufacturing performances compared to liquid electrolytes. However, the biggest obstacle impeding their practical applications is the limited ionic conductivity, which is often several orders of magnitude lower than commercialized liquid electrolytes \cite{Song2023, Zhou2019, Lopez2019}. Extensive experimental \cite{Alarco2004, Bouchet2013, Zhang2018, Pesko2016} and computational research \cite{Savoie2017, Webb2015, xie2022accelerating} have been undertaken to tackle this challenge. Common strategies to improve the ion conductance in polymers encompass operating at high temperatures \cite{Steele1983}, adding secondary additives \cite{Tang2012, Bandara1998}, copolymerizing\cite{bouchet2013single, C3CC49588D}, and formulating polymer composite electrolytes \cite{Liu2017, Wang2021}. Nonetheless, these approaches predominantly rely on poly(ethylene oxide) (PEO) \cite{Fenton1973} which is known as the most widely used SPE material, whereas there has been relatively limited exploration of alternative non-PEO polymers. In addition, PEO possesses limited performance at room temperature because of its tendency of crystallization \cite{Varzi2016}. Given the vast molecular space upon which polymers build and the intricate hierarchical nature of polymer networks, substantial untapped potential exists at the bottom for designing new SPEs with enhanced ionic conductivity. However, the vast polymer space also poses the challenge of efficiently searching for promising candidates for specific applications in a brute-force way.

To address the challenge, machine learning (ML) and data-driven approaches have been widely used for property prediction \cite{Pilania2019, Tran2020, Khajeh2023} and inverse design \cite{Barnett2023, Kim2021, Chen2020} with desirable functions of polymers, an area known as polymer informatics. One essential step in the polymer informatics pipeline is translating the polymer’s compositional and structural information into data representation. Typically, a single polymer chain's representation is either adapted from molecular codes like simplified molecular-input line-entry system (SMILES) \cite{Weininger1988, Lin2019}, molecular graph \cite{Queen2023} of repeat unit, or expressed through descriptive numerical values, often referred to as fingerprints or descriptors\cite{Rogers2010, Kuenneth2023}. While ML algorithms and platforms used for other fields in materials informatics should be equally applicable to polymer informatics, a notable challenge is the scarcity of accessible databases, especially those with reliable property labels of diverse polymers. Therefore, in recent years, an increasing number of studies have concentrated on augmenting the polymer dataset pool through literature mining \cite{Otsuka2011}, hypothetical polymerization \cite{Kuenneth2023}, high-throughput molecular dynamics (MD) simulations \cite{xie2022accelerating, 10.1063/5.0160937}, and generative design \cite{Ma2020}. The development of these polymer databases will vastly facilitate the advancement of polymer informatics. 

Among the various ML approaches, generative design distinguishes itself due to its ability to learn from data distribution and generate novel candidates. This paradigm holds promise not only for expanding the polymer database by learning from existing data but also for crafting polymer materials tailored to specific design objectives. Generative models, such as variational autoencoders (VAEs) \cite{Bombarelli2018}, generative adversarial networks (GANs)\cite{de2018molgan}, and recurrent neural networks (RNNs)\cite{Kotsias2020}, have already been extensively employed for small molecule generation. These techniques have been further adapted to the field of polymer informatics, driving the generation of polymers with target properties such as excellent thermal conductivity \cite{Ma2022}, great gas permeability \cite{Yang2023}, exceptional dielectric performance \cite{Gurnani2021}, and high synthetic viability \cite{Kim2023}. With the recent advances in generative AI, such as transformer-based architectures and diffusion models, new opportunities and more potent solutions are offered to address the challenges of polymer generation.  

\begin{figure}[h]
    \centering
    \includegraphics[width=0.5\textwidth]{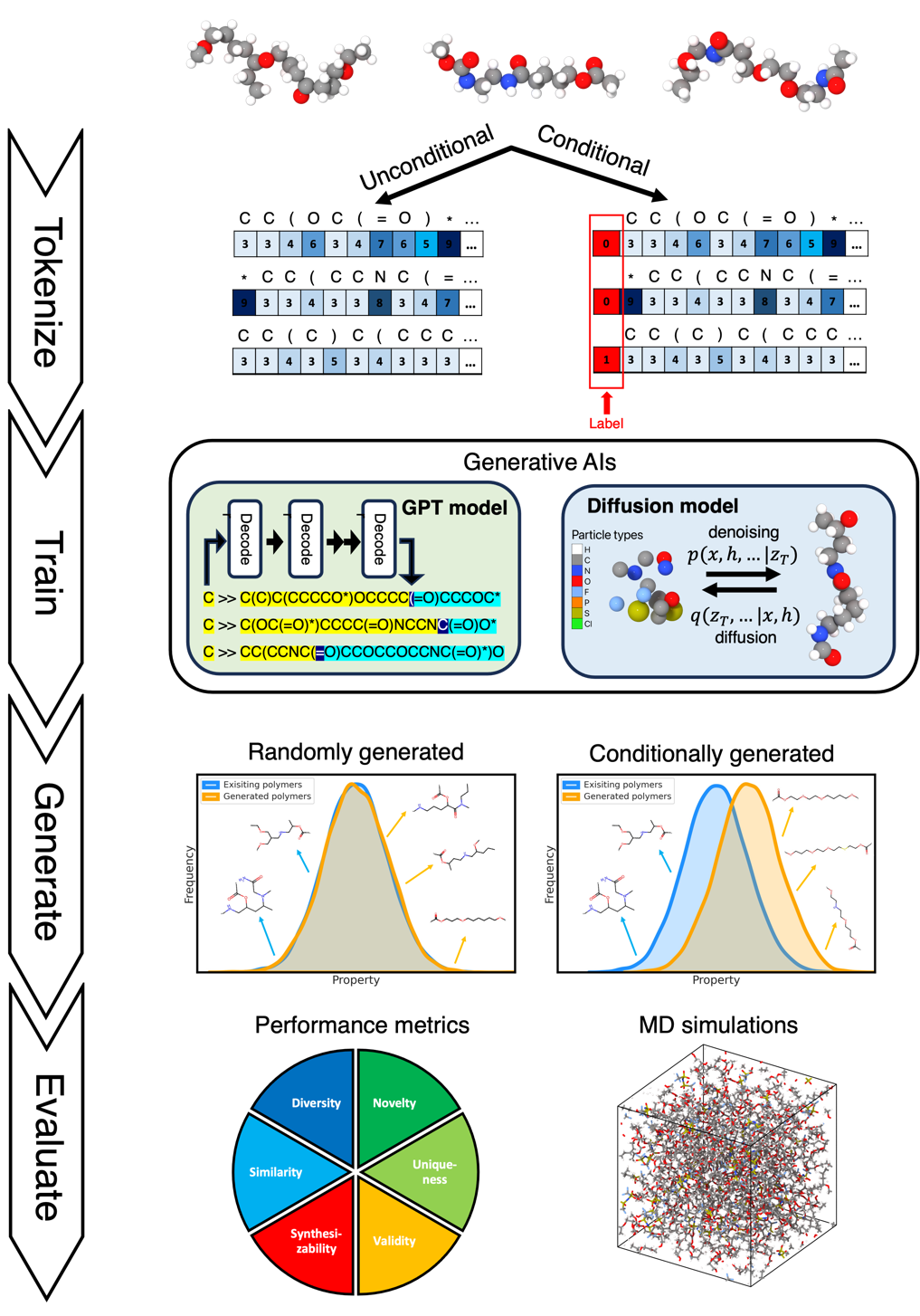}
    \caption{\textbf{Depiction of the workflow of the work reported here.} Starting from HTP-MD dataset, we tokenize the SMILES codes which represent the monomers of polymer electrolytes. Both unconditional generation (label-free) and conditional generation (conductivity class label inserted in the sequence) are performed with GPT-based and diffusion-based generative models. The generated polymer set is evaluated with different metrics, and promising candidates with high conductivity from conditional generation are further validated with molecular dynamics simulations.}
    \label{fig:1}
\end{figure}

In this study, we delve deep into the intricacies of different advanced generative models for polymer generation. To train the generative models, we leverage a previously curated dataset (HTP-MD dataset) consisting of 6024 different amorphous polymer electrolytes with their ion transport properties computed from MD simulations \cite{10.1063/5.0160937, xie2022accelerating}. In our current investigation, we compare the performance of different generative models’ architectures in (Fig. \ref{fig:1}): 1) Unconditional generation, in which the model is trained to learn the “language” of polymers, enabling it to produce novel, chemically valid, unique, and synthesizable polymers. 2) Conditional generation, in which we aim to steer the generative model towards generating candidates exhibiting high ionic conductivity along with other metrics defined for unconditional generation. To ensure high-quality polymer generation, we compare two categories of advanced generative models: one based on a minimal implementation of a large language model (LLM) known as generative pretraining transformer (GPT) and the other based on diffusion models evaluated with multiple metrics. Hyperparameter optimization is undertaken to optimize the performances of these generative models. Through our proposed methodology, we can easily produce a large number of monomer units for polymer electrolytes, which can significantly enrich the current dataset. Furthermore, the top 45 candidates from conditional generation with high ionic conductivity are validated using full-atom MD simulations. Among these candidates, we've identified 17 polymer units whose ionic conductivity surpassed all existing data in our trainset, and multiple polymers increased the optimal conductivity value in the train set by more than 100\%. Finally, given the limited data amount and highly constrained polymer space of the HTP-MD dataset, we apply a “pretraining \& fine-tuning” strategy to enhance the model’s performance \cite{Ma2020}. By employing this strategy, the model attains accelerated convergence and improved data adaptability. In a concurrent study~\cite{Lei2023}, we introduce an advanced polymer discovery framework that leverages the top-performing model from this research. The framework enhances polymer identification through an active feedback loop which augments the dataset on the fly with generated high-performance polymers validated by MD simulations. This self-improving framework further allows the fast discovery of novel polymer electrolytes with significantly improved ionic conductivity.

\section{Results}

\subsection{Data preparation, model selection, and evaluation}\label{subsec2}

To convert the chemical representations of polymers into a format comprehensible to AI, we employed the polymer SMILES string (p-SMILES). This string is an adaptation of the standard SMILES notation for chemicals, combined with special characters “*” to encode ends of the monomer unit in a homopolymer chain (e.g. p-SMILES of PEO “*OCC*”). The p-SMILES string is further tokenized at the character level, analogous to preprocessing human text, which is used as input for training language models. The p-SMILES strings are padded to ensure uniform length for all monomer units of polymers within the HTP-MD dataset (see Methods section for more details). 

We here examined multiple recently developed advanced generative AI approaches for polymer generation, including a GPT-based model and two diffusion-based models. GPT model is a state-of-the-art DL model developed by OpenAI for natural language processing and understanding\cite{Brown2020}. The model is grounded in the Transformer architecture\cite{Vaswani2017}, renowned for its self-attention mechanism that empowers the model to dynamically weigh and focus on relevant input tokens, thus efficiently capturing long-range dependencies within a sequence. Moreover, while many attention-based models are bidirectional or encoder-only, GPT exclusively uses a decoder-based architecture, emphasizing the generation aspect of its design. Considering the amount of data and the feasible training effort, we employ a minimal version of the GPT model\cite{Karpathy2022}, referred to as “minGPT” in subsequent discussions.

 Diffusion models are a class of deep generative models that iteratively refine the generation over a series of steps, simulating a process akin to the diffusion phenomenon in physical systems. This process involves invertible transitions between random noises to the target data distribution. The forward process, termed the diffusion step, incrementally introduces noise to the original input. Conversely, the reverse process, labeled the denoising step, counteracts the forward process, reconstructing data samples from the noises. 

In this study, we employ two variants of diffusion-based models. The first is a 1D denoising diffusion probabilistic model\cite{Ho2020}, hereafter termed "1Ddiffusion," and the second is a diffusion language model~\cite{Li2022}, referred to as "diffusion-LM" in subsequent discussions. The primary distinction between these two diffusion models lies in the space where the diffusion and denoising actions take place (Fig. S1). In the 1Ddiffusion model, noise is both introduced and eliminated directly within the token space, meaning the token index changes throughout the process. By contrast, the diffusion-LM model begins by encoding discrete tokens into continuous embeddings (“embedding” step) and then learns to introduce and remove noise within this embedding space. Finally, a learnable reverse step (“rounding” step) decodes the embeddings back into the original token space in the diffusion-LM model. More detailed information about both GPT-based and diffusion-based generative models is included in the Methods section. 

Given that three models possess distinct loss functions and that loss values may not provide a comprehensive evaluation for chemical systems, we here propose an evaluation method for polymer generation with six different metrics. These six metrics are novelty, uniqueness, validity, synthesizability, similarity, and diversity:
\begin{enumerate}
   \item Novelty: measures the proportion of generated polymers that do not exist in the training set, as the generation target is producing unseen polymers.
   \item Uniqueness: gauges the percentage of non-duplicate polymers within the generated set to evaluate whether the model can generate different polymers.
   \item Validity: calculates the proportion of chemically valid SMILES codes to prove that the model learns the correct chemical language. 
   \item Synthesizability: assesses the percentage of polymers considered easy to synthesize for potential experimental synthesis in the future.
   \item Similarity: offers a similarity score comparing both the composition and structure of polymers in the training set with those in the generated set, indicating whether the model is learning from the existing polymers. 
   \item Diversity: provides a dissimilarity score among the generated polymers, evaluating whether the generated polymers are diverse so that no mode collapse issue is witnessed in the generation. 
For all the metrics, a higher value indicates better performance. These metrics or similar metrics have been utilized in various prior studies for evaluating molecule generation tasks\cite{Bilodeau2022}. A detailed quantitative breakdown of how these metrics are computed can be found in the Methods section. 
\end{enumerate}

In order to compare different models and optimize their performances, we first perform a grid search for hyperparameter optimization. For each model, we adjust three independent hyperparameters to assess the model's performance across different hyperparameter combinations. For a detailed breakdown of the chosen hyperparameters and their respective values, please refer to the Methods section. Here, we utilize the average value from six metrics to identify the optimal hyperparameter combination, given that all metrics are within the same range and scale. 

\subsection{Unconditional generation}

\begin{figure}[h]
    \centering
    \includegraphics[width=0.7\textwidth]{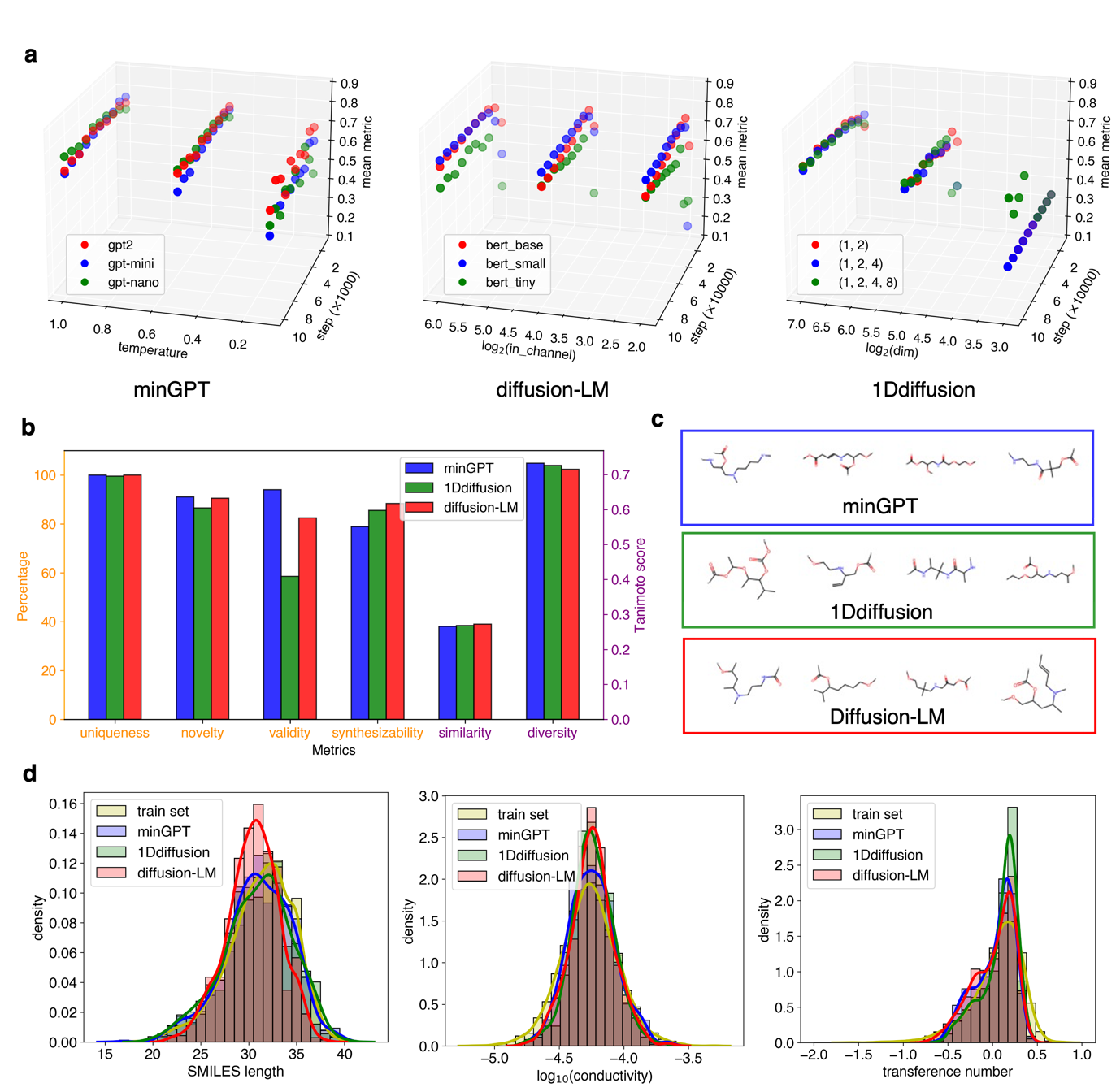}
    \caption{\textbf{Unconditional generation and model comparison.} a) Comparison between average performances of models in grid search of hyperparameters tuning. b) Comparison between the performances of different models in each evaluation metric. c) Example monomer structures generated by minGPT, diffusion-LM, and 1Ddiffusion models during unconditional generation task. d) Reproduction of property distributions comparing train set and generated set.}
    \label{fig:2}
\end{figure}

We first trained generative models to learn the polymer language without property constraints. This endeavor seeks to establish a foundation for pretraining on unlabeled dataset investigated in the subsequent sections and create a universal model for polymer electrolytes that is not limited to one or two specific objectives. The performances of minGPT, 1Ddiffusion, and diffusion-LM models under different hyperparameter combinations are displayed in Fig. 2a. All the metrics are normalized between 0 and 1. As the figure indicates, the optimal performance of minGPT (mean metric = 0.773) and diffusion-LM models (mean metric = 0.767) are comparable, while the 1Ddiffusion model (mean metric = 0.736) reaches a relatively lower best score compared to the other two. We further plot all six metrics for an optimal model of each kind in Fig. 2b. For both the minGPT and diffusion-LM models, we observe high levels of novelty, uniqueness, validity, and synthesizability, with most values exceeding 0.8. This suggests that the majority of the polymers generated are novel, valid, and potentially synthesizable. In contrast, the 1Ddiffusion model underperforms compared to the other two models in producing valid polymers. Example polymer structures generated by those models are shown in Fig.2c. The stars in the molecular structures indicate the two ends of polymers which are connected repeatedly later to form a longer polymer chain with about 150 atoms in the backbone for MD simulation validation. 

In addition, we further compare the performances of these models by assessing their ability to replicate distributions of various polymer properties, including the length of the SMILES string, conductivity, and transference number. Among these properties, the transference number is the fraction of the total ionic charge carried by a particular ion species in the electrolyte. In this work, we refer to the transference number specifically as the cation transference number. As for polymer electrolytes, a higher ionic conductivity and cation transference number will lead to more efficient and safer battery operation\cite{https://doi.org/10.1002/advs.202003675}. Given these properties, we examined how well the distribution of the generated set aligns with the training set (Fig. 2d). To calculate the ionic conductivity and transference number, we utilized a graph neural network (GNN) which was developed in our previous study\cite{10.1063/5.0160937}. The GNN model was also trained on the HTP-MD dataset and outperformed classical ML algorithms like random forest. As shown in the figures, the minGPT model (blue lines in figures) reproduces the distribution accurately for all three properties. In contrast, the other two diffusion-based models (green \& red lines) tend to produce narrower distributions for various properties compared to the training data. The quantitative evaluation is based on the difference calculation of two distributions using Kullback-Leibler (KL) divergence.

Finally, in terms of computational costs, the minGPT model is more efficient in both training and inference compared to the diffusion-based models. Training the optimal minGPT model requires around 3-4 mins on a Tesla V100 GPU core (16GB RAM), while both optimal 1Ddiffusion and diffusion-LM models take about 2 hours to train. Regarding inference, the diffusion models, due to their step-by-step denoising from random input, also have a higher computational overhead for polymer generation. This becomes especially important when generating thousands of polymer candidates. In conclusion, when evaluating the three models based on metric assessments, property distribution replication, and computational efficiency, the minGPT model clearly shows the best performance. It surpasses both the 1Ddiffusion and diffusion-LM models by delivering superior performance across all three criteria. Detailed scores of all metrics are listed in Table S1. 

\subsection{Conditional generation}

\begin{figure}[h]
    \centering
    \includegraphics[width=0.7\textwidth]{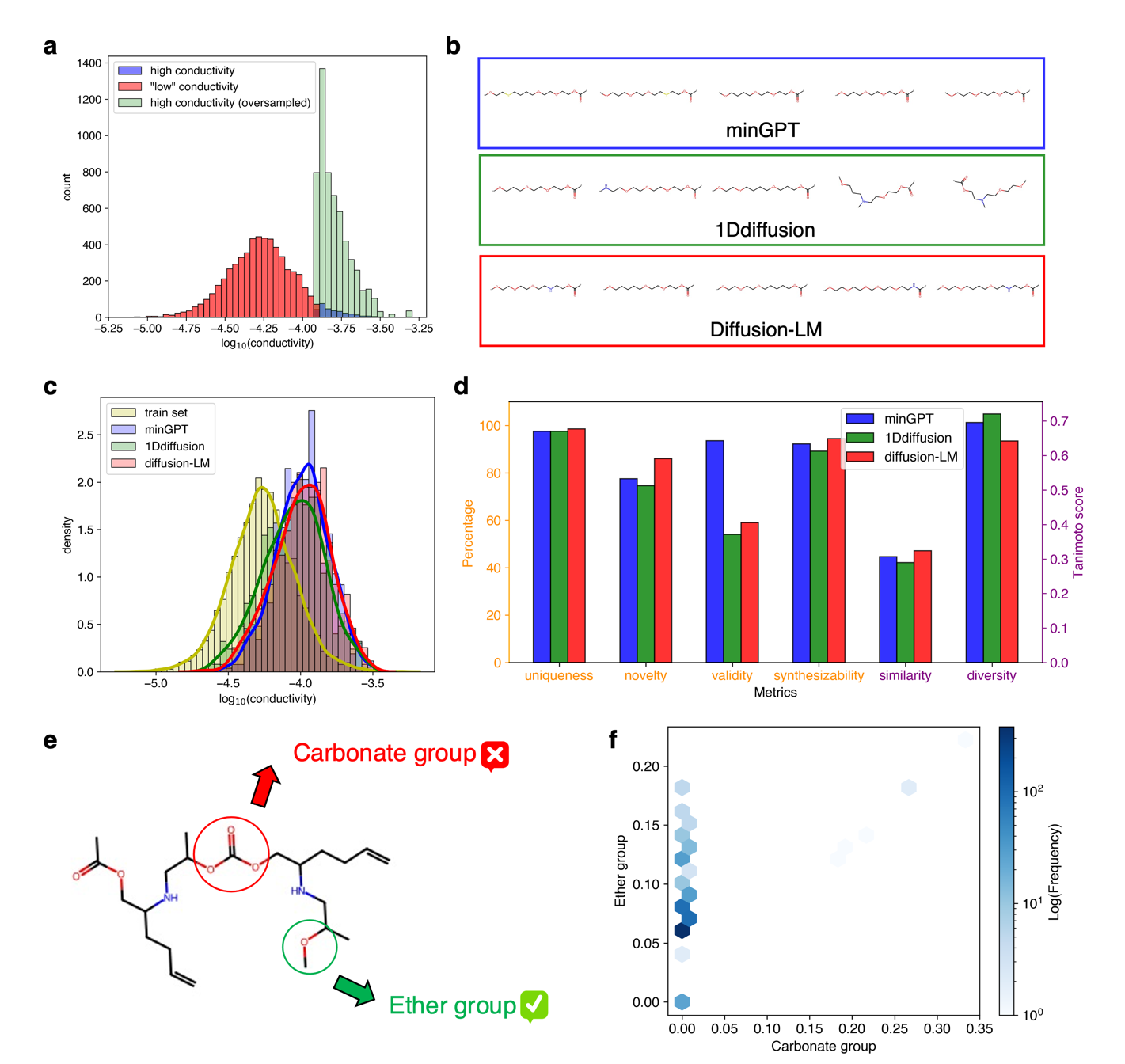}
    \caption{\textbf{Conditional generation on ionic conductivity and functional group. } a) Oversampling to create a balanced dataset. Blue and green histograms are the distributions of high ion conductive polymers before and after oversampling. b) Example monomer structures generated by 3 models during conditional generation task. c) Shift of conductivity distributions comparing train set and generated set. d) Evaluation metrics of model performance. e) Conditional generation with target functional groups. The carbonate group is an unwanted functional group, while the ether group is a desirable one. f) Frequency of functional groups with conditional constraints. The value of each axis represents the fraction of ether and carbonate groups among all functional groups inside the monomer. }
    \label{fig:3}
\end{figure}

Beyond unconditional generation, we further aim to steer the generative models towards creating polymer electrolytes with desirable properties such as high ionic conductivity. To achieve this objective, we first categorize polymers from the HTP-MD dataset into high-conductivity and low-conductivity groups. Specifically, the top 5\% of polymer electrolytes are selected as the high-conductivity group, which results in an imbalanced dataset. To prevent training a biased model, we oversample the high-conductivity polymers by randomly replicating the data. After oversampling, the number of polymers in the high-conductivity group is the same as that of the low-conductivity group with the data distribution as depicted in Fig. 3a. The conductivity label is incorporated into the input sequence using special characters (“[Ag]”, “[Ac]”) that are not present in any of polymer sequences (see Methods section for more information). Therefore, generative models like the GPT-based model can initiate with conductivity labels and subsequently generate highly ionic conductivity polymer units. In terms of diffusion-based models, our approach simply involves training them to generate polymer sequences with conductivity labels and then selecting sequences marked with high-conductivity tokens. 

We utilize the optimal model architectures obtained from hyperparameter tuning of unconditional generation tasks to produce polymer electrolytes with high conductivities. The GNN model proposed in our previous study\cite{10.1063/5.0160937} is employed to predict the ionic conductivity of generated polymers. Fig. 3c displays the conductivity distribution, comparing the generated set with the training set, which reveals a distinct shift toward the high-conductivity domain for all generated sets from the three models. Once again, we plot the model comparison using the six metrics (Fig. 3d). Aligning with the findings from the unconditional generation, the minGPT model surpasses both the 1Ddiffusion and diffusion-LM models, achieving a superior average score. The top five candidates from the generated sets of the three models are presented in Fig. 3b. In contrast to the diverse structures of unconditionally generated monomer units, the conditionally generated polymers with the highest conductivities exhibit consistent patterns (Fig. 3b). For example, majority of the generated polymers possess linear backbones and contain the “-O-CH2-CH2-“ fragment within their chains. Given the high ionic conductivity of PEO-like polymers, such an observation is to be expected.

Beyond conditioning polymer generation on conductivity, we can also design polymer electrolytes with specific functional groups. By strategically designing these groups, we can offer valuable guidance for easier experimental synthesis, such as proposing potential precursors and reaction pathways. In addition, this approach allows us to incorporate specific functional groups based on domain knowledge or avoid those that result in undesirable properties for polymer electrolytes. As an illustration, we present the design of polymers containing ether groups (R-O-R’) while excluding carbonate groups (R-OC(=O)O-R’) as shown in Fig. 3e. Ether groups will potentially enhance ion conductance in polymer electrolytes as the oxygen atoms contain negative charges that can contribute to ion diffusion process\cite{Xue2015}. In contrast, polymers with carbonate groups are likely to form a liquid-like phase because the carbonate group is readily cleaved, especially at high temperatures, and no longer useful as solid polymer electrolytes\cite{Davis1969}. To design polymer electrolytes with specific functional groups while excluding undesired ones, we include two labels to the polymer sequence: one representing the ether group and the other for the carbonate group. Fig. 3f shows the results of conditional generation on functional groups. As the figure manifests, most of the generated polymers do not contain any carbonate groups, while more than half of the polymers in the training set contain carbonate groups, yet they all possess the ether group with varying numbers between instances. The results of polymer generation conditioning on conductivity and functional groups show that our method can be applied to diverse design objectives. 

\subsection{Molecular dynamics validation}

\begin{figure}[h]
    \centering
    \includegraphics[width=0.7\textwidth]{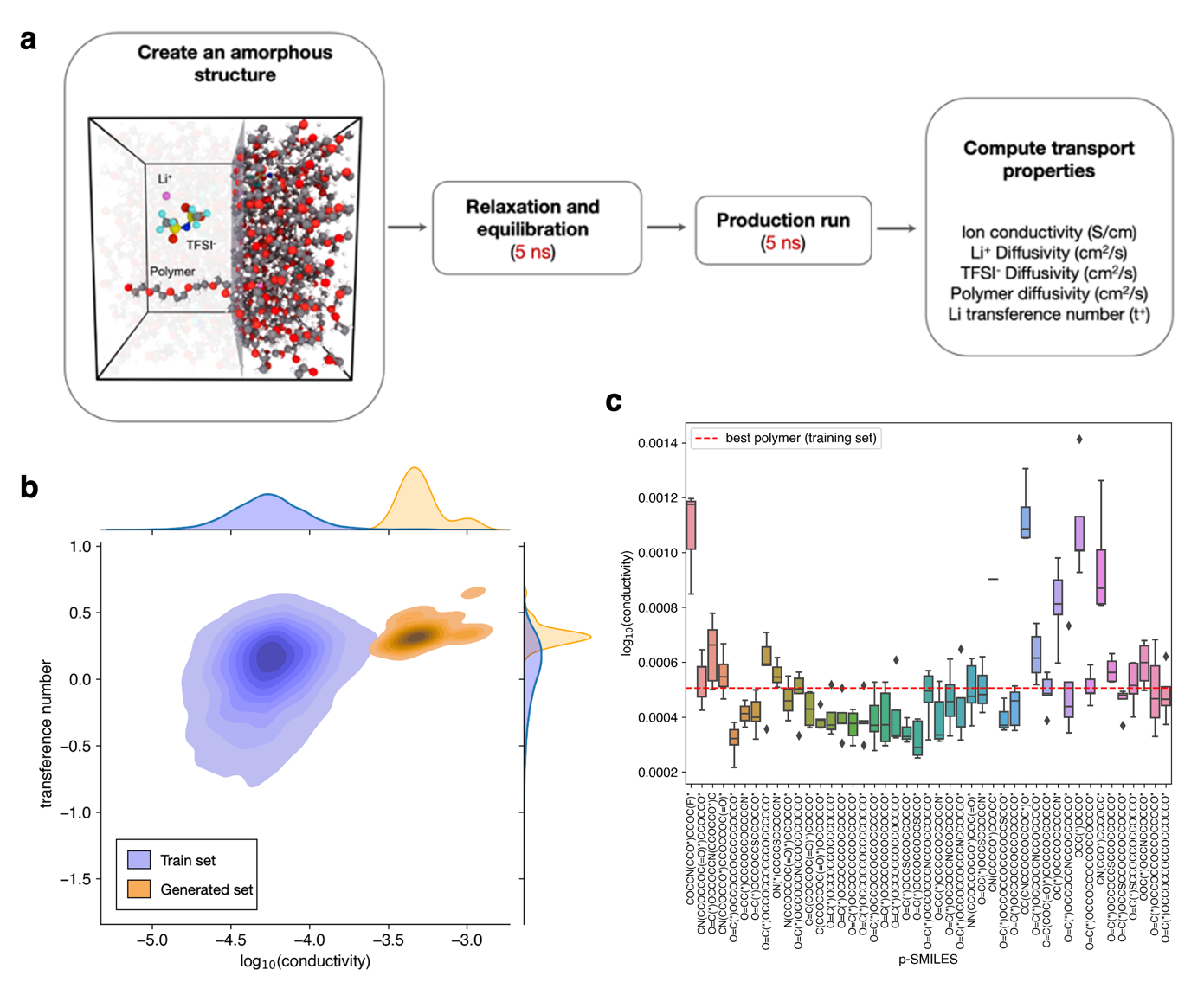}
    \caption{\textbf{MD validation. } a) Schematic illustrating the MD simulation workflow. The process begins with the creation of an amorphous polymer-salt system. Subsequent steps include system relaxation and equilibration, followed by the production run. Ion transport properties are then determined using the cluster Nernst-Einstein method based on data collected during the production phase. b) Distribution of conductivity and transference number calculated using MD simulations for the top 45 candidates in the generated set (100K samples in total) in comparison to the training data.  c) List of conductivity values for each candidate. There are 17 out of 45 candidates showing better conductivity than the optimal polymer (conductivity = 5.1×10-4 S/cm) in the train set.  }
    \label{fig:4}
\end{figure}

With the generative model for conditional generation, we further create 100,000 polymers with high-conductivity labels. The larger size of this generated set increases the likelihood of identifying highly conductive candidates. To evaluate the conductivity of these 100K candidates, we again utilized the previously developed GNN model. Based on this assessment, the top 50 candidates were selected for further validation through MD simulations. Our simulation approach aligns with established high-throughput MD methodologies applied in previous polymer electrolyte studies~\cite{xie2022accelerating, Khajeh2023}. The process initiates with the formation of an amorphous polymer-salt system, followed by steps of system relaxation, equilibration, and a production run (Fig. 4a). Ion transport properties are subsequently determined using the cluster Nernst-Einstein method, leveraging data from the production phase. More details about the simulation procedure can be found in the Methods section.  However, due to stability issues and limitations of the force field, only 45 out of the 50 candidates were successfully simulated. 

Fig. 4b presents the distribution of ionic conductivity and transference number of the top 45 candidates among the 100K generated candidates. As indicated in the figure, the ionic conductivities of these promising candidates are significantly higher than those in the existing training dataset. Notably, 17 of the 45 candidates surpass the highest conductivity polymer (conductivity = 5.07×10-4 S/cm) in the training set (Fig. 4c). The best-performing candidate in the generated set whose p-SMILES code is CC(CNCCOCCOCCOC*)O* achieves a conductivity of 1.13×10-3 S/cm, more than double that of the best polymer in the training set. The complete list of these top 45 candidates, along with their ionic conductivities and transference numbers, can be found in Table S2. Interestingly, despite not specifically targeting this property during conditional generation, the transference numbers of the generated polymers are generally higher than those in the training set. This suggests a possible positive correlation between ionic conductivity and transference number. 

With MD simulations as a validation tool, a closed-loop framework can be established. Each iteration within this framework comprises a sequence of steps: conditional generation of polymers, preliminary screening via our ML predictions, evaluation through MD simulations, and data augmentation incorporating new findings. In a separate study~\cite{Lei2023}, we have explored this opportunity by presenting a robust polymer discovery framework based on improving generative models’ performance through iteratively incorporating MD simulation feedback. We have shown this approach can facilitate the generation of promising polymer electrolyte candidates originating from the refinement of trained machine learning models. We envision integrating this approach with high-throughput synthesis methodologies in the future to pave the way for the development of a self-driving laboratory.  

\subsection{Interpretability}
Indeed, the generative models we utilize, like many other deep learning models, are often criticized for their black-box nature, which offer limited physical or scientific insights. To better understand why the generated polymers possess high conductivity, we analyze the occurrence frequency of functional groups among the top candidates.  For a more robust statistical analysis, we expand our focus from the top 45 to the top 1000 candidates out of the 100K generated to analyze various functional groups. These include ketone, carbonate, ester, ether, secondary/tertiary amine groups, and sp3 hybridized carbon atoms (denoted as “AlkylCarbon”). 

In our analysis of functional groups, we observed a notable increase in the frequency of the ether group, as illustrated in Fig. S2. An increased number of ether groups contributes to higher conductivity, as these groups contain negatively charged oxygen atoms. These atoms are actively involved in the ion diffusion or hopping processes, thereby significantly enhancing ionic conductivity\cite{Munshi1988, Zhao2020}. In contrast, while the ketone group also contains an oxygen atom, its double bond structure is more rigid. This rigidity, coupled with the introduction of side chains by the ketone group, leads to a less flexible polymer backbone. Conversely, alkyl carbon atoms contribute to easier deformation of polymer backbone with greater flexibility, resulting in a lower viscosity. As a result, ion transport in these polymers is less restricted, leading to increased conductivity\cite{Ye2010}. This is why we witness fewer ketone groups, and more sp3 hybridized carbon atoms in the top candidates (Fig. S2). The analysis of functional group frequencies here not only provides interpretability to the results generated by our model but also offers promising design principles for actual experimental synthesis. 

\subsection{Pretraining \& fine-tuning}
Given the limited number of data within the HTP-MD dataset, we further employ a pretraining \& and fine-tuning strategy to exploit the capacity of generative AIs. These two phases have demonstrated a significant enhancement in the performance of LLMs. During pretraining, the objective is to harness all available data sources, enabling the model to attain a broad understanding of language. Subsequent fine-tuning with task-specific datasets refines and tailors this generalized model to specialized domains or tasks. 

Since the minGPT model surpasses diffusion-based models in both unconditional and conditional generation, we employ this model to evaluate the impact of the pretraining and fine-tuning approach. Initially, minGPT is pretrained on the PI1M dataset, a benchmark database comprising one million monomer units of polymers\cite{Ma2020}. The polymers within the PI1M dataset are generated using a model trained on the PolyInfo database\cite{Otsuka2011}, which aggregates data from academic literature.
The PI1M dataset lacks information regarding the ionic conductivity of the polymers. As such, we utilize the minGPT model for unconditional generation during the pretraining phase. Then, the pretrained minGPT model is fine-tuned to generate polymers with high conductivity, as previously described.  

\begin{figure}[h]
    \centering
    \includegraphics[width=1.0\textwidth]{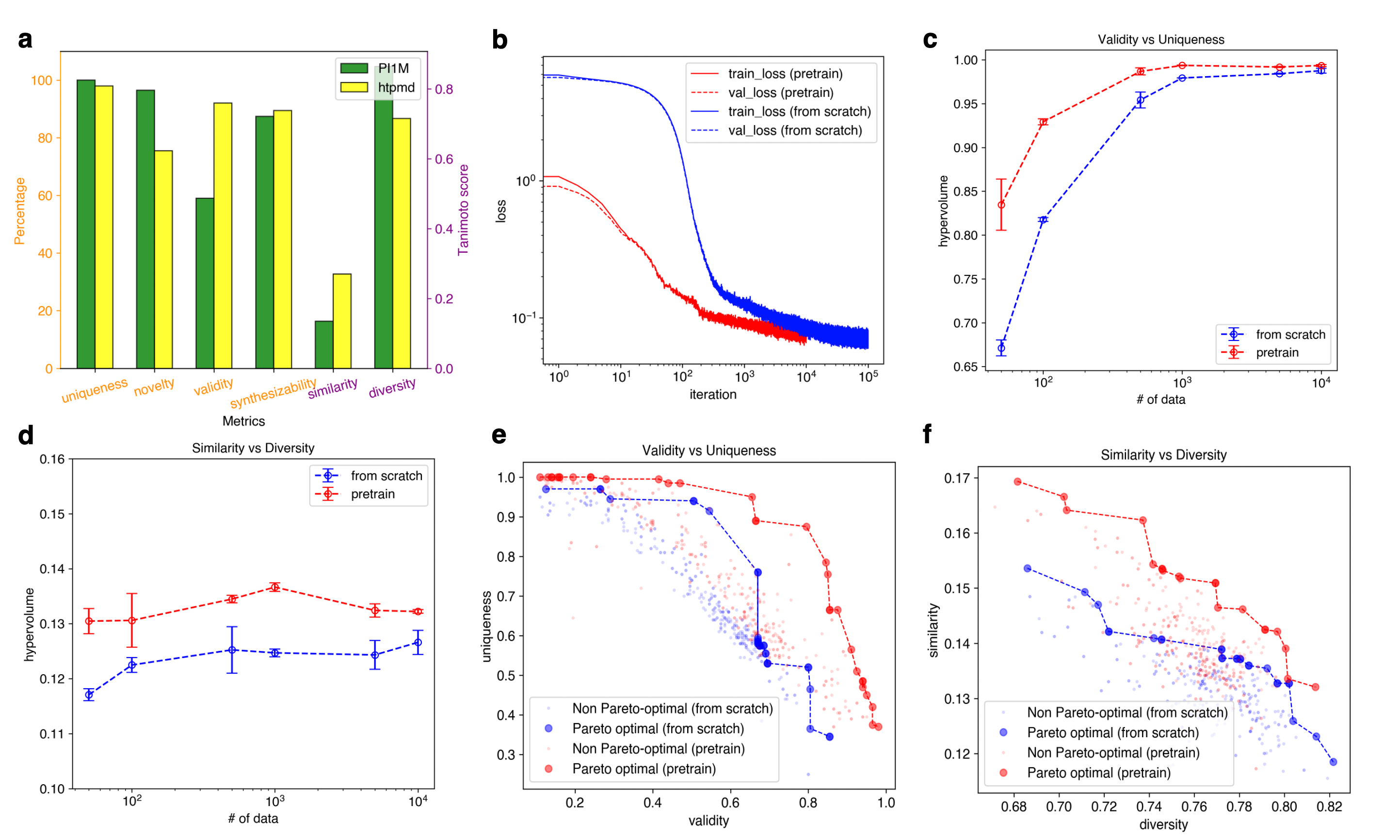}
    \caption{\textbf{Pretraining \& fine-tuning. } a) Performances of models pretrained on PI1M and HTP-MD datasets. b) Loss curves comparing pretraining + fine-tuning and training from scratch. c) Model performance with varying numbers of data (hypervolume calculated with respect to validity \& uniqueness). d) Model performance with varying numbers of data (hypervolume calculated with respect to similarity \& diversity). Each point in the plot corresponds to a different set of hyperparameters. e) Pareto front comparing pretraining + fine-tuning and training from scratch (validity \& uniqueness). f) Pareto front comparing pretraining + fine-tuning and training from scratch (similarity \& diversity).}
    \label{fig:5}
\end{figure}

We compared two minGPT models: one trained on the HTP-MD dataset (from scratch) and the other pretrained on the PI1M database for unconditional generation and then fine-tuned on HTP-MD dataset to conditionally generate polymers (Fig. 5a). The PI1M dataset encompasses a broader range of polymer structures, including aromatic rings not found in the HTP-MD dataset. Consequently, the model exhibits greater diversity and reduced similarity scores compared to the version trained on the HTP-MD dataset from scratch. However, this increased structural diversity also posed challenges in ensuring the chemical validity of the generated polymers, leading to a decreased validity score for the model pretrained on the PI1M dataset.

Through the implementation of pretraining and fine-tuning, we observed three notable improvements in the model's performance by comparing it with the model trained solely on HTP-MD dataset. Above all, pretraining shortened the training time for fine-tuning (Fig. 5b). While some might argue that pretraining is a more time-consuming phase, it's essential to note that this label-free pretrained model, which learns the general language of polymers can serve as a foundational model for a multitude of downstream tasks, not just the conditional generation based on ionic conductivity. 

Secondly, pretraining improves the validity and uniqueness scores of conditional generation, especially when only a small amount of data is available for fine-tuning (Fig. 5c). This improvement is crucial, given that when training with experimental data, we often face a scarcity of such data due to the prohibitive costs and extended durations associated with high-throughput experiments. We here primarily showcase the results of validity and uniqueness scores as synthesizability is highly correlated with validity, and novelty increases monotonically as uniqueness increases (Fig. S3). Therefore, it is sufficient to utilize validity and uniqueness scores to illustrate the advantages of using a pretrained model. We vary the number of training data and evaluate models’ performances under different combinations of hyperparameters as we did in the unconditional generation section. Given that there is an inherent trade-off between these two metrics (Fig. 5e, each dot in the figure corresponds to a different set of hyperparameters), we utilize the hypervolume in a multi-objective context as our evaluation metric. Further discussions regarding the origin of this trade-off and the rationale behind our metric selection can be found in the Methods section. As demonstrated in Fig. 5c, the performance of models trained on the entire HTP-MD dataset is similar, regardless of whether pretraining was conducted. However, as the dataset size diminishes, the difference in performance between models with and without pretraining becomes increasingly significant, underscoring the value of pretraining when fine-tuning with a limited amount of data. The Pareto fronts (Fig. 5e) of validity versus uniqueness at a small data number (50 data only) further proves the model with pretraining shows a consistently better performance. 

Finally, we reveal that pretraining enhances the model’s capacity to capture the polymer characteristics of fine-tuning dataset (HTP-MD dataset), while also producing a diverse range of polymers during the generation, as evidenced by improved similarity and diversity metrics. The improvement is always evident regardless of the number of training data (Fig. 5d). The diverse polymer structures and rich chemical information in the PI1M dataset underpin the superior performance achieved through pretraining. Again, the Pareto fronts are visualized (Fig. 5f), showing that the optimal solutions from the model with pretraining consistently outperform those from the model without pertaining. 

\section{Discussion and Conclusion}
In this study, we compare and benchmark advanced deep generative AI models to learn the complex language of polymers, enabling the generation of promising electrolyte candidates. Our proposed methodology significantly enriches the relatively undeveloped data pool of polymer electrolytes. We test two mainstream generative models including GPT and diffusion-based models. The GPT-based models outperform diffusion-based architectures with better performance and lower training/inference costs. The results of hyperparameter tuning indicate that small models can also show great performances given the limited number of data we have. We design a systematic evaluation scheme for polymer generation with multiple metrics, and leverage it to comprehesively evaluate the generation tasks. Given the positive and negative correlations between these metrics, we further examine the model performance in a multi-objective setup.  Moreover, by conditioning the generative process, our models are capable of producing polymers tailored to specific target properties. Empowered by our proposed approach, the generation of novel, unique, chemically valid, and potentially synthesizable polymers can be accomplished in microseconds. This significantly surpasses the capabilities of human chemists, offering a dramatic acceleration in the materials discovery process. By applying this method, we have already discovered multiple polymer candidates that surpass all existing data from our HTP-MD dataset. 

By integrating MD simulations as a validation tool, we have established a targeted polymer discovery framework with comprehensive closed-loop workflow in a parallel study~\cite{Lei2023}. This workflow will encompass a series of sequential steps: generative design of new polymers, calculating properties of generated polymer candidates via MD simulations which serves as a validation technique, and the enrichment of our database with these new findings. This iterative loop is capable of continuously generating new electrolyte candidates and progressively refining the developed models. Furthermore, the scope of this workflow extends beyond electrolyte material discovery. Considering the vast array of applications for polymer materials, this method shows tremendous potential in revolutionizing various polymer systems. It could significantly impact sectors such as biomedical devices, sustainable packaging solutions, advanced aerospace components, and high-performance textiles, among others. Each of these fields could benefit from the enhanced efficiency and targeted material properties this workflow promises. 

Another contribution of this study is showing how LLMs and other large deep generative models can be utilized for the generative design of polymers under conditions of limited labeled data availability. The pivotal process in this context is pretraining, which effectively maximizes the potential of these models and enables them to learn from broader contexts before being fine-tuned for specific applications. Although the capacity of LLMs is not fully exploited when we train the model with our small HTP-MD dataset, the implementation of LLMs can really benefit the generation when pretrained on a vast array of diverse polymers. This pretraining and fine-tuning strategy empowers large foundational models to surpass traditional generative models like GANs or VAEs, in domain-specific and design-focused tasks. Moreover, the ability of these models, even with sparse datasets, is crucial, especially when applying this workflow to experimental data, which is often scarce, unstructured, and expensive to generate.

Looking ahead, our current polymer candidates are poised for experimental synthesis and testing. Simultaneously, our approach can continuously generate and evaluate new polymer candidates, supplying them for experimental examination. Leveraging the comprehensive evaluation metrics proposed in this work and the self-refining framework, we believe that the candidates developed through our computational methods hold tremendous potential for actual synthesis and demonstrating high conductivity in practical experiments. In addition, we can also further integrate the experimental feedback into refining our workflow. This could involve introducing additional constraints during the conditional generation phase. An illustrative example of this is presented in the conditional generation part of our Results section. Based on insights from our experimental collaborators, we learned that polycarbonate is not stable in their high-throughput experimental setup. Consequently, we adapted our polymer generation process by excluding carbonate groups, as previously discussed. Another method to incorporate experimental insights is to directly apply our proposed workflow to experimental data. This approach could pave the way for the development of a self-driving laboratory, where new test cases are intelligently generated using our generative models. Such a system would not only streamline the discovery process but also enhance the precision and relevance of the test polymers generated. 

Finally, this work only touches on the minimal level of complexity of polymer electrolyte materials, focusing on the chemistry of a single repeating unit. However, the true complexity of polymers lies beyond. For instance, polymers can exhibit heterogeneous compositions within a single chain (e.g. copolymers). With multiple polymer chains, they can mix, blend and entangle to form high-level intricate polymer networks. These compositional and structural variations play a crucial role in determining the electrochemical properties of polymers.  Moreover, inside a battery, there are salts and other additives that actively interact with polymer electrolytes. The combination of these materials strongly affects the overall performance of the battery. While the ultrahigh complexity and vast range of possible combinations pose great challenges in designing these materials, they also unveil immense potential. We are confident that our approach holds great potential for adaptation to complicated polymer design problems and facilitates the search for next-generation battery materials. 

\section{Methods}
\subsection{p-SMILES representation and preprocessing}
In this work, homopolymer repeat units are denoted by p-SMILES strings. p-SMILES follows the same syntax definition as SMILES~\cite{Lin2019} but uses two stars to indicate the two endpoints of the monomer unit. The p-SMILES strings are further tokenized into sequential representation using the DeepChem package\cite{Ramsundar2019}. All sequences are padded to a uniform length of 64. Considering the 6024 distinct repeat units of polymers within the HTP-MD dataset, there are 20 total tokens, which include special characters that signify the start and end of a sequence as well as a padding token. We split the data into training and test sets, with 80\% of the data used for training and 20\% for testing. In terms of pretraining and fine-tuning strategy, we use the whole SMILES character list from DeepChem to tokenize both PI1M and HTP-MD datasets, as they contain different numbers of tokens. 

In the case of unconditional generation, a random integer between 1 and 10 serves as the initial token for minGPT to start next-token prediction. Meanwhile, diffusion-based models generate sequences directly from Gaussian noises that match the input's length. To signify high-conductivity and low-conductivity categories, respectively, we use two additional special characters, “[Ag]” and “[Ac]” (referred to as the "conductivity token" in the main text), which are not presented in any of the polymer sequences. To reinforce this constraint, we replicate the conductivity token, forming a sequence of 5 tokens in length, before inserting it at the beginning of the polymer sequence for the generative task using minGPT. As for diffusion-based models, a single conductivity token inside the sequence suffices for conditional generation. 

\subsection{Model architectures}
The code used for the minGPT model is adapted from open-source codes available on the Github repository karpathy/minGPT~\cite{Karpathy2022}. The model first translates the sequence of tokens into two input embeddings, including token embedding and positional embedding. This positional embedding is essential to counteract the innate lack of sequence awareness characteristic of Transformers. The embeddings are then passed through several layers of Transformer blocks, each boasting a multi-head self-attention mechanism paired with a feed-forward neural network. Within these blocks, layer normalization and residual connections are interwoven, providing stability to activations (GELU activation function~\cite{Hendrycks2016}) and aiding in the training of deeper networks. A unique facet of GPT is its approach to parameter sharing across all Transformer layers, which stands in contrast to other Transformer-based models. 

For diffusion models, they define a Markov chain composed of diffusion steps that introduce noise to the original data. Subsequently, the model is trained to reverse this process, constructing data from random noises. It's crucial to highlight that only the denoising step is trainable; the diffusion process is deterministic once the hyperparameters is decided. Considering the limited volume of the paper, we here skip related detailed mathematical derivation and layer-by-layer explanation of model architectures but focus on highlighting the key components of the models used in this work. 

Among the two diffusion models we use for polymer generation, the 1D diffusion model is a modification of the denoising diffusion probabilistic model (DDPM)~\cite{Ho2020}, initially created for image generation. A distinguishing feature of DDPM is its use of Gaussian noise in the forward diffusion phase, facilitating a closed-form sampling technique known as reparameterization trick. This means we can directly sample data at any arbitrary timestep without accumulatively calculating through each individual diffusion step. Furthermore, by converting the 2D UNet architecture to 1D—primarily through substituting 2D convolutional layers with 1D layers—we've adapted it for sequence generation. The token space is discrete, with tokens identified by distinct integers, while the diffusion model operates on continuous space. Therefore, we apply a rounding operation to transform continuous values from the denoising process into discrete token indices. The code used in this work is based on the open-source Github repository lucidrain/denoising-diffusion-pytorch~\cite{Wang2023}. 

In the diffusion-LM model, a non-autoregressive language model architecture is merged with a continuous diffusion concept for text generation. The texts, in our case, p-SMILES strings, are first encoded with an embedding function that maps each token to a vector in continuous embedding space parameterized by a Gaussian distribution. Besides the token embedding and positional encoding commonly found in Transformer-based language models, the diffusion-LM model also incorporates additional timestep embedding, which encodes information related to the diffusion process. In the inverse process, the trainable rounding function essentially identifies the most probable token at each position based on the learned probability distribution, analogous to multi-label classification. To minimize errors stemming from the rounding function, loss reparameterization~\cite{Kingma2013, Rezende2014}, and clamping tricks~\cite{Li2022} are subsequently applied. The general architecture of the diffusion-LM model is based on well-established LLMs like bidirectional encoder representation from Transformers (BERT)~\cite{Devlin2019}. The codes utilized for the diffusion-LM model in this study are modified based on the open-sourced Github code XiangLi1999/Diffusion-LM~\cite{Li2022-repo}. 

\subsection{Hyperparameter selection}
As mentioned in the main text, we select three independent hyperparameters for each model. One common hyperparameter for all three models is the training step. The rationale for investigating the training step as a hyperparameter is its influence on model performance: longer training typically results in decreased loss, increased validity and synthesizability, yet it can diminish novelty and uniqueness. This presents a trade-off between the metrics of novelty/uniqueness and validity/synthesizability (more in-depth discussions are available in the “Pretraining \& fine-tuning” section). Hence, adjusting the training step is crucial to navigate the Pareto front of these metrics. 

Apart from the training step, other hyperparameters are used for different models and both selected and optimal values of these hyperparameters are listed in Table S3. The other two hyperparameters of the \texttt{minGPT} model are \texttt{model architecture} and \texttt{temperature}. The ``model architecture'' corresponds to the HuggingFace’s Transformer-based model architecture~\cite{Wolf2019}. Meanwhile, ``temperature,'' a characteristic feature of GPT-style models usually ranging from 0 to 1, dictates the degree of randomness incorporated into the generation process. Notably, a higher temperature leads to more random, diverse, and creative text generation, but it also increases the likelihood of producing incoherent or irrelevant outputs, often termed ``hallucinations.'' As for the \texttt{1Ddiffusion} model, \texttt{dim} is the initial hidden dimension of 1D UNet and \texttt{dim\_mult} determines the scaling of layer dimensions. For instance, with \texttt{dim\_mult} set to (1, 2), the hidden dimension of the second layer is twice that of the first layer. Finally, when it comes to diffusion-LM models, \texttt{model architecture} is again the HuggingFace’s model. The \texttt{in\_channel} is the input channel dimensions of embeddings. The detailed results for the hyperparameter grid search for three models are listed in Tables S4, S5, and S6. To summarize the findings presented in the tables, the impact of enhancing the model's complexity, either through deeper architecture or more intricate hyperparameter configurations, parallels the effect of increasing training steps. Such modifications typically enhance the model's validity and its ability to synthesize data. However, they may also lead to a reduction in novelty and uniqueness. 

\subsection{Evaluation protocol}
In this section, we delve deeper into the six evaluation metrics employed in our study. As outlined in the main text, "novelty" quantifies the proportion of polymers that are unseen relative to the training set, while "uniqueness" measures the proportion of non-duplicate polymers within the generated set. These metrics are essential because, during the generation process, varying input noises can lead to identical polymer outputs, given the discrete nature of polymers. 

“Validity” which is defined based on the chemical validity of SMILES strings is evaluated using RDKit package~\cite{landrum2016rdkit}. There are mainly two additional principles for a p-SMILES string to be chemically valid besides being a valid SMILES string: 1) It must have two endpoints (two stars) that are connected. 2) The bond connecting these two endpoints must be a single bond.  

Regarding "synthesizability," this metric is derived from the Schuffenhauer's Synthetic Accessibility (SA) score previously developed for evaluating the synthesizability of drug-like molecules~\cite{Ertl2009}. The SA score theoretically evaluates a molecule's ease of synthesis by considering both fragment contributions and a complexity penalty~\cite{Ertl2009}. This score ranges from 1 (indicating easy synthesis) to 10 (signifying high difficulty). In our approach, polymers with an SA score below 5 are considered synthesizable\cite{Ma2022}. It's important to recognize that variations in the specific threshold for the SA score do not substantially influence the overall evaluation of synthesizability. We can also utilize the mean SA score which shows the same trend as the synthesizability as defined.

When it comes to “similarity” and “diversity”, both metrics are built on so-called Tanimoto similarity~\cite{Bajusz2015} which evaluates the chemical similarity of two molecules by comparing the correlation between their molecular fingerprints like Morgan fingerprint~\cite{Morgan1965}. The Tanimoto score has a range between 0 and 1, where a higher score indicates greater similarity. The similarity score is calculated using the average score of all paired polymers between the training and generated sets, which is employed to determine if the generative model captures the distribution of the existing data. Furthermore, it's essential that the generated polymers display a diverse range of structures, ensuring that the model doesn't suffer from the 'mode collapse' issue often encountered in generative models like GANs (e.g. all generated human face images are similar to each other)~\cite{Creswell2018}. As a consequence, the diversity score is quantified by determining the average dissimilarity (1 – similarity score) among all paired polymers within the generated set. In summary, by evaluating both similarity and diversity, we ensure that the generated polymers not only align with patterns in the existing data but also exhibit distinct differences from one another. 

To evaluate models’ performances under massive different combinations of hyperparameters during the grid search, 200 polymers are generated for comparison in each case given the interest of time and cost. This allows us to identify the most promising model. Once we've selected the optimal model, we then produce a larger sample set with 1000 polymers to thoroughly assess its final performance. 

\subsection{Functional group analysis}
To analyze the frequency of functional groups within polymers, we first polymerize the monomer unit into a dimer to ensure that the functional groups at the connecting part of the repeated units can also be captured. Once the dimer is built, we utilize the Identify Functional Group (IFG) python package~\cite{Riddle2021} to find available functional groups and their occurrences to calculate the frequency. The frequency is defined as the number of functional groups over the number of atoms in the dimer of the polymer. We only visualize the main functional groups with occurrence frequency higher than 0.01. As a result, functional groups, including ketone, carbonate, ester, ether, secondary/tertiary amine groups, and sp3 hybridized carbon atoms are discussed. 

\subsection{MD simulations}
We performed MD simulations on polymer-(Li+.TFSI-) systems at 353 K with a salt molality of 1.5 mol/kg in Large Atomic Molecular Massively Parallel Simulator (LAMMPS)\cite{PLIMPTON19951}, using Polymer Consistent Forcefield (PCFF+)\cite{https://doi.org/10.1002/jcc.540150708}.  The simulation procedure included initial relaxation and equilibration of the polymer-salt mixture, followed by a production run to gather data for calculating ion transport characteristics, such as ionic conductivity. The equilibration phase comprised sequential NVT (constant number of particles, volume, and temperature) and NPT (constant number of particles, pressure, and temperature) ensembles to systematically adjust the system to densities approximating theoretical predictions. The NVT simulation at 353 K was followed during the production run for 5 ns with a time step of 2.0 fs. Finally, the trajectories are analyzed and ion transport properties are computed via cluster Nernts-Einstein equitation~\cite{PhysRevLett.122.136001}. More information on MD simulation can be found elsewhere\cite{10.1063/5.0160937, xie2022accelerating}.   

\subsection{Multi-objective problem}
To help understand the origin of the trade-off between validity/synthesizability and uniqueness/novelty, here is an analogy in the context of LLM: A general LLM or chatbot (high uniqueness) is capable of many tasks but master of none, leading to potential hallucinations in specific domain knowledge (low validity). Conversely, an LLM fine-tuned for a specialized purpose (high validity) is like a master craftsman in one domain, but may lack the broad versatility of the generalist (low uniqueness). 

Given the trade-off between these metrics, the evaluation of model performance is thus a multi-objective problem. In a multi-objective problem, there is no single solution that is uniquely optimal given that the increase of one objective will lead to decrease of others. Instead, there is a set of optimal solutions which forms so-called Pareto front, where no solution is strictly better than another in all objectives. We vary the hyperparameters, as in previous sections, to examine the Pareto front for models both with and without pretraining. The "hypervolume" is a metric that quantifies the "size" of the space dominated by the Pareto front. It represents the volume of the region in the objective space that is dominated by the set of solutions on the Pareto front. Essentially, a larger hypervolume indicates a Pareto front that dominates a more substantial portion of the objective space, suggesting a more optimal set of solutions. Thus, the metric can be utilized to evaluate the models’ performance. For each data set size, we conduct the hyperparameter grid search three times to determine the average values and standard deviations of the hypervolumes across various models, as shown in Fig. 5c and Fig. 5d.

\section*{Data availability}
All data are available in the main text and Supplementary Information. Additional data that support the findings of this work are available from the corresponding author upon reasonable request.

\section*{Code availability}
The source codes are deposited in Github:  https://github.com/TRI-AMDD/PolyGen

\section*{Acknowledgments}
We extend our deepest gratitude to Professor Grossman and Professor Yang Shao-Horn, along with Dr. Tian Xie, Dr. Sheng Gong, and Dr. Arthur France-Lanord at the Massachusetts Institute of Technology for their invaluable support in establishing our molecular dynamics simulations workflow. Their guidance has been instrumental in advancing our research.
Our sincere thanks also go to Professor Jeremiah Johnson for engaging with us in enlightening discussions about the synthesizability and stability of polymer electrolytes under experimental conditions. His insights have significantly contributed to the robustness of our study.
Additionally, we appreciate the thought-provoking dialogues with Professor Rafael Gomez Bombarelli regarding generative models, which have enriched our understanding and application of these complex systems.
The collaboration and input from these esteemed colleagues have been a cornerstone of our work, and we are thankful for the expertise and knowledge they have shared.

\section{Disclosures}

The authors wish to acknowledge that the method used to inversely design polymers using our generative model, including tokenization and incorporation of the properties of polymers to the input as described within this manuscript, is subject to a provisional patent application. This application has been submitted with the following details: U.S. Patent Application No. 63/582,871 titled "Methods of Designing Polymers and Polymers Designed Therefrom" with TEMA Reference No. IP-A-6823PROV and Darrow Reference No. TRI-1107-PR. The authors confirm this does not alter our adherence to arXiv policies on sharing data and materials.

\section*{Author contributions}
Z.Y., W.Y., X.L., H.-K.K., D.S., and A.K. conceived the idea. Z.Y., with A.K., designed the experiments. Z.Y., W.Y., X.L., and A.K. were involved in the development of generative models for polymers. Z.Y. executed the experiments, while A.K. performed the molecular dynamics (MD) simulations. Data analysis and visualization were carried out by Z.Y. and A.K. All authors, Z.Y., W.Y., X.L., H.-K.K., D.S., and A.K., were actively involved in writing, reviewing, and editing the manuscript at various stages.

\bibliographystyle{unsrt}  
\bibliography{references}  

\end{document}